# Transmodal Analysis of Neural Signals

Yaroslav O. Halchenko[a,*], Michael Hanke[a,b], James V. Haxby[a], Stephen José Hanson[c], Christoph S. Herrmann[d]

[a]*Department of Psychological and Brain Sciences, Dartmouth College, Hanover, NH 03755, USA*
[b]*Department of Psychology, University of Magdeburg, Magdeburg, Germany*
[c]*Psychology Department, Rutgers, Newark, NJ 07102, USA*
[d]*Department of Psychology, Carl von Ossietzky University Oldenburg, Oldenburg, Germany*

**Abstract**

Localizing neuronal activity in the brain, both in time and in space, is a central challenge to advance the understanding of brain function. Because of the inability of any single neuroimaging techniques to cover all aspects at once, there is a growing interest to combine signals from multiple modalities in order to benefit from the advantages of each acquisition method. Due to the complexity and unknown parameterization of any suggested *complete* model of BOLD response in functional magnetic resonance imaging (fMRI), the development of a reliable ultimate *fusion* approach remains difficult. But besides the primary goal of superior temporal and spatial resolution, conjoint analysis of data from multiple imaging modalities can alternatively be used to segregate neural information from physiological and acquisition noise. In this paper we suggest a novel methodology which relies on constructing a quantifiable mapping of data from one modality (electroencephalography; EEG) into another (fMRI), called transmodal analysis of neural signals (TRANSfusion). TRANSfusion attempts to map neural data embedded within the EEG signal into its reflection in fMRI data. Assessing the mapping performance on unseen data allows to localize brain areas where a significant portion of the signal could be reliably reconstructed, hence the areas neural activity of which is reflected in both EEG and fMRI data. Consecutive analysis of the learnt model allows to localize areas associated with specific frequency bands of EEG, or areas functionally related (connected or coherent) to any given EEG sensor. We demonstrate the performance of TRANSfusion on artificial and real data from an auditory experiment. We further speculate on possible alternative uses: cross-modal data filtering and EEG-driven interpolation of fMRI signals to obtain arbitrarily high temporal sampling of BOLD.

## 1. Introduction

Due to complementary characteristics of EEG[1] and fMRI acquisition methods, there is an increasing number of reported EEG/fMRI conjoint studies, which attempt to gain the advantages of multiple imaging modalities in experiments involving perceptual and cognitive processes. Researchers who approach multi-modal data collection and analysis run into a multitude of problems regarding the acquisition (Laufs et al., 2007) and analysis of neural data (see Halchenko et al., 2005; Herrmann and Debener, 2007, for an overview of existing studies and analysis methods).

---

[*]Corresponding author
  *Email address:* yaroslav.o.halchenko@onerussian.com (Yaroslav O. Halchenko)
  *URL:* http://www.onerussian.com (Yaroslav O. Halchenko)
  [1]Although suggested methodology is applicable to MEG as well, only EEG allows for in-magnet acquisition, and is therefore investigated in the paper.

The main obstacle on the way to multimodal analysis is the absence of a complete model of the BOLD[2] response in fMRI. A number of models have been suggested: On one hand they include naive modeling of BOLD signal in the context of a Linear Time Invariant System (LTIS) (Friston et al., 1994) using a convolution of experimental design variables or features of neural activation (*e.g.*, power of EEG bands) with a predefined hemodynamic response function (HRF). On the other hand there are more comprehensive models of the BOLD signal in terms of detailed biophysical processes (*e.g.*, *Balloon* (Buxton and Frank, 1997) or *Vein and Capillary* (Seiyama et al., 2004) models). It was shown in multiple studies, that the BOLD signal is an inherently non-linear function of neuronal activation (Boas et al., 2008) that varies across subjects (Aguirre et al., 1998) and brain regions (Handwerker et al., 2004). Therefore, former simpler models are not general enough to explain the variability of the BOLD signal, whereas complex parametric models rely heavily on a prior knowledge of nuisance parameters (due to biophysical details) for which there are no reliable and straightforward means of estimation.

Despite an inherent non-linearity of the BOLD response, there have been multiple reports of detecting a linear dependency between the observed BOLD response and local field potentials (LFP) (Logothetis and Wandell, 2004, and references 27, 29, 54, 55 and 81 therein). In general, such results are not inconsistent with the non-linearity of BOLD, since a non-linear function can be approximately linear in the context of a specific experimental design, region of interest, or dynamic ranges of the selected features of EEG signals. *Sufficient* linearity inspired multimodal fusion attempts aiming at the ultimate goal of superior spatio-temporal resolution, where unknown parameters of a forward BOLD model could be estimated via Bayesian inference under heavy prior assumptions to provide the required regularization of the solution (*e.g.*, Daunizeau et al., 2007). Generic application of such methods to real EEG/fMRI data is hindered by the necessity of choosing priors proven to be trustworthy for such generative modeling. Because of that difficulty most of current fullbrain multimodal analysis strategies either rely on preconditioning of EEG inverse solutions with tentative activation loci obtained from fMRI data analysis (*e.g.*, Dale and Sereno, 1993; Liu et al., 1998; Wagner and Fuchs, 2001), or exploring fMRI data for the areas (or some spatial components such as ones obtained from spatial ICA analysis (*e.g.*, Eichele et al., 2007)) with significant covariates to features of EEG signals (*e.g.*, N170 component in Horovitz et al., 2004), or a reasonable description based on a convolutional model with some arbitrary HRF such as in LeVan et al. (2009). In such approaches, analysis relies upon canonical or deduced from sparse event-related design shape of HRF. "Hunting for covariates" approach allowed to detect areas where significant portions of BOLD signal variance could be explained by linear relationship to the power of EEG frequency bands (such as $\alpha$-rhythm (de Munck et al., 2007), Rolandic $\alpha$- and $\beta$- EEG rhythms (Ritter et al., 2009) *etc.*). Further studies have shown that high-frequency bands ($f \geq 100\,\text{Hz}$) carry explanational power as well (Martuzzi et al., 2008). Most of the analyses considered a single band of the EEG data from a particular electrode at once and relied on an arbitrarily chosen HRF model to provide necessary temporal smearing of the neural effect to mimic desired BOLD time-course. Only recently Zumer et al. (2009) approached linear correlational multimodal analysis between multiple MEG frequency bands and BOLD signal to show that more of BOLD variance could be explained if multiple MEG bands were considered at once. Moreover, considering flexible models of HRF has been shown (van Houdt et al., 2010) to provide better detection power over relying on some canonical HRF.

To summarize, independent researchers found various EEG signal features that can explain significant portions of the BOLD response variance. Moreover, considering multiple frequency bands at a time and relaxing assumptions over the HRF further improved the description of the empirical BOLD signal. Based on these findings, we suggest a methodology for multimodal *fusion* analysis – a transmodal analysis of neural signal (called TRANSfusion), which relies on constructing a *reliable model* to describe BOLD signal in terms of EEG data. Reliability is assured by the fact that the constructed model is not simply a best-fit but rather a generalizable model of BOLD signal in terms of the time-varying spectral representation of EEG from multiple frequency bands and electrodes at the same time. In comparison to simple correlational analyses, actual modeling and validation of the mapping go beyond the initial goal of detection where relation between features of EEG and BOLD could be observed. Furthermore, in TRANSfusion we do not use any specific (conventional or estimated) model of hemodynamic response, and rather rely on constructing a generalizable forward model of BOLD response per each voxel.

---

[2]Although other protocols are available for fMRI, blood-oxygenation level dependent (BOLD) protocols are most popular. Therefore, fMRI and BOLD terms are used interchangeably throughout the article.

Subsequent analysis of BOLD reconstruction accuracy on data unseen during model estimation allows us to reliably localize brain regions which were activated during the experiment, and therefor contributed to the signal acquired in both neural data modalities. Further analysis of the model provides spatial relevance profiles for any particular EEG feature used as mapping input (*e.g.*, specific frequency band, or a particular input channel). This apporach is similar to the analysis of classifiers sensitivity for fMRI and EEG data which was shown to be more powerful than respective univariate methods (Parra et al., 2002; Hanson et al., 2004; Hanson and Halchenko, 2008; Hanke et al., 2009).

Here we describe TRANSfusion analysis methodology, verify its validity on simulated data, and demonstrate its applicability to real data from auditory EEG/fMRI experiment. On the latter dataset we show that TRANSfusion does not only confirm results of conventional univariate analysis of fMRI data, but also localizes additional activation loci. A consecutive analysis allows to associate loci with specific components of EEG signal, such as $\alpha$-rhythm or functional relation of areas to signal obtained at an EEG electrode of interest.

## 2. Methods and Data

### 2.1. Formulation of the Approach

The main goal of TRANSfusion is to learn a mapping function $\mathcal{F}$ to provide a reliable description of BOLD signal in terms of EEG data, such that

$$y_i = \mathcal{F}_i(\mathbf{X}), \tag{1}$$

where an fMRI signal $y$ at each voxel $i$ is described in terms of EEG data $\mathbf{X}$ from multiple channels via a transformation $\mathcal{F}_i$.

We believe, that because of loose or absent coupling between the power of some frequency bands and the design of the experiment, it is desirable to analyze the data acquired simultaneously from both modalities in a single session. Due to impossibility of acquiring MEG in the magnet, TRANSfusion should be especially beneficial for the conjoint analysis of EEG and fMRI. The idea of deriving a per-voxel mapping $\mathcal{F}$ is similar to general linear model (GLM) based analysis of fMRI data where each voxel is represented as a linear combination of a limited number of explanatory variables. However, it substantially differs from GLM analysis in that it is data driven and it neither imposes a specific shape of hemodynamic response function (HRF), nor depends on a specific experimental design. Moreover, it does not strictly require linear relationship between $\mathbf{X}$ and $y_i$ and could be readily extended to non-linear mappings.

*Richness* of EEG signals (*e.g.*, in terms of its spectral properties), absence of a generative forward model of BOLD signal, and sluggishness of BOLD response are the main obstacles toward deriving a reliable transformation function $\mathcal{F}$. The slow evolution of BOLD response implies that a vast duration of EEG signal preceding any given volume is required to provide an adequate description for BOLD signal. To accommodate for the lag between neural activation and its assumed reflection within the fMRI signal, (1) can be further elaborated as

$$y_i(t) = \mathcal{F}_i(\mathbf{X}(t - \tau \dots t)), \tag{2}$$

where $t$ is a time onset of fMRI volume acquisition (fMRI volumes are usually evenly spaced in time with a fixed repetition time (TR), typically 2–4 sec), and $\tau$ is a reasonable duration to retroactively account for neural activations which could be significant contributors to the fMRI response at the given onset $t$. Such setup is schematically depicted in Figure 1. In most of the scenarios, it is reasonable to take $\tau = 10\,\text{sec} + \text{TR}$. A duration of 10 sec should be sufficient to account for the typical HRF delay of BOLD response. The duration of TR is included since different slices of a volume are acquired at different offsets within the TR. Hence for an atypically long TR (as it is in the experimental data of Section 2.4, where TR=10 sec) it is necessary to consider sufficient amount of EEG data to describe any voxel in the acquired volume.

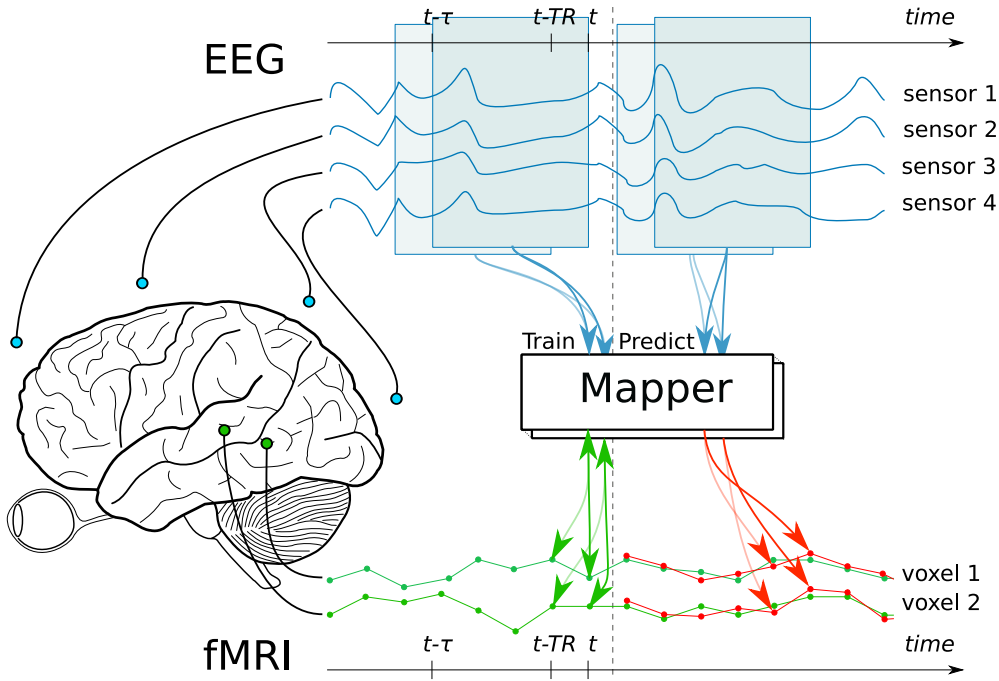

Figure 1: Mapping of neural data from EEG into fMRI. In training phase a single mapper (regression) is trained on a corresponding time-window of EEG data to predict fMRI signal in a single voxel of fMRI. During testing the trained mapper predicts fMRI signal. Comparison between predicted (in red) and real time course of fMRI (in green) allows to assess the performance of the mapper and to state either given voxel carries a signal which could be described by EEG.

Additionally, it is possible to account for the offset of any given slice while formulating (2), if slice ordering is known. But such procedure might be inappropriate since fMRI data are conventionally motion corrected as a part of preprocessing. Motion correction relies on the interpolation between different slices, hence it can *smear* the effect of slice acquisition timing.

Due to the interest in using the power of various spectral EEG bands as the input to the model, transformation $\mathcal{F}$ could be described as a composition of actual regression function $\mathcal{G}$ and some time-frequency transformation $\mathcal{T}$ (*e.g.*, SFFT, Wavelet standard or packet decomposition)

$$y_i(t) = \mathcal{G}_i(\mathcal{T}(\mathbf{X}(t-\tau\ldots t))), \tag{3}$$

where $\mathcal{T}$ does not depend on the voxel in question and is introduced to convert the original EEG data into a feature space which is known to have correspondence with BOLD response. Transformation $\mathcal{T}$ can also incorporate multiple characteristics of the same feature of the EEG signal, *e.g.*, for a single time-frequency slot it could utilize both amplitude and the power of the EEG signal. That allows to account for some non-linear effects which were reported for the BOLD response without requiring transformation $\mathcal{G}$ itself to be non-linear.

Having selected features of EEG data to be involved in the mapping, many electromagnetic source imaging methods could naturally be extended to account for fMRI data if a generative forward model of BOLD signal was available. For instance, direct universal-approximator inverse methods (Jun and Pearlmutter, 2005) have been found to be very effective (fast, robust to noise and to complex forward models) for the EEG dipole localization problem, and could be augmented to accept fMRI data if the generative model for it was provided.

Although there are no explicit constraints on a method to estimate regression function $\mathcal{G}$ in (3), due to abundance of number of input variables (*i.e.*, multiple sensors, time-frequency representation for $\tau$) some multivariate regression methods could not be chosen. Naive regressions could heavily overfit the data, and thus lack any generalization ability (*i.e.*, not to be representative for the data in general). Therefore, we resorted to a method developed in statistical learning: Support Vector $\epsilon$-insensitive Regression (SVR). SVR is inherently regularized, and as a result, is capable to provide reasonable generalization performance even if the dimensionality of input space greatly exceeds the number of available data samples (*e.g.*, Poldrack et al., 2009). Linear SVR has simple parametrization (just the trade-off coefficient $C$ with a reasonable default set to the inverse of the mean norm of the samples) and efficient software implementations. Besides having great generalization performance on high-dimensional data, supervised learning methods encourage model testing (*e.g.*, via cross-validation) to actually provide unbiased estimates of the generalization performance for a derived transformation $\mathcal{F}$.

Furthermore, many learning methods conveniently provide estimates of sensitivity (or feature-weighting) of the constructed mapping, measuring the impact of each feature on the prediction or even on the generalization ability of the learner (*e.g.*, Rakotomamonjy, 2003). Such sensitivity assessment is quite banal for linear learners if data was normalized across features prior training. Furthermore it can be estimated for non-linear systems via noise-perturbation (*e.g.*, Hanson et al., 2004) or can be derived for many classes of non-linear learners with theoretical generalization bounds, (*e.g.*, SVM; Rakotomamonjy, 2003). Such sensitivities can be used to determine which components of the signal were found to be relevant for a given regression or classification task, thus providing means for localization (*e.g.*, Hanke et al., 2009).

## 2.2. Computational Efficiency

Since TRANSfusion scans a brain volume in a mass-univariate fashion by estimating transformation function $\mathcal{G}_i$ per each target voxel, it seems to demand an extreme amount of computational resources, since typical fMRI volume has up to 200,000 brain voxels (the actual number of voxels depends on the spatial resolution of the data). However, in the suggested approach each input sample for the regression is the same multivariate set of features of EEG data, *i.e.*, input is invariant across all voxels. The target time-course of an fMRI voxel is the only variable, different per each voxel, thus the model of hemodynamic response to be trained per each voxel. Because of invariant input data, most kernel-based learning methods (such as SVR, Gaussian Process Regression (GPR), *etc.*), need to compute the kernel matrix only once for a specific set of parameters. Pre-computing of kernel matrices on the full dataset (prior cross-validation) and selection of required kernel rows and columns for a given cross-validation split only, completely obviates kernel computation insided the cross-validation loop (which has to be carried out per each voxel). Initial optimization point for the iterative methods (such as constrained quadratic optimization in SVR) can also be pre-seeded with a previously found solution for another (neighboring) voxel. Due to the guaranteed uniqueness of the solution, such initialization of optimization might result in faster convergence, and must result in nearly exact (up to a specified numerical tolerance) solution as if it was done with full training and arbitrary initial starting point for optimization. Aforementioned "shortcuts" allow to perform learning of the model per each voxel in a reasonable amount of time (*e.g.*, from ten to hundreds of milliseconds on a regular PC), therefore making whole brain analysis feasible. Due to the mass-univariate nature, estimation could also be parallelized for the computation in high performance computing environments, theoretically making real-time processing of the incoming data possible.

## 2.3. Simulated Data

Any novel methodology has to be validated first on data with known characteristics of noise and the signal of interest (*i.e.*, of spatio-temporal signals of the neuronal activation in case of neuroimaging). Due to the absence of a realistic study of the activations in a phantom, it was necessary to simulate the signal and noise conditions.

The simulated environment consisted of a single relevant slice within a three shell (brain, skull, scalp) spherical model of the head (see Figure 2(a)). fMRI resolution was set to 3 mm (isotropic). At this resolution the modeled "brain" consisted of 2148 voxels. Temporal resolutions for the simulated signals were in line with those of the empirical dataset (analysis of which will be presented in Section 2.4): 50 Hz for EEG, and 0.2 Hz (equivalent to TR=5 sec) for fMRI[3]. Three regions of interest (ROIs), colored in red, dark red, and blue in Figure 2(a), were defined to carry event related (ER) activity, whereas all other locations carried only spontaneous activity, amount of which depended on the tissue type (gray or white). For EEG modeling, each voxel contained a single dipole, with its orientation set to the normal of the artificial folding (depicted in yellow) of the cortical tissue. Eight out of 16 EEG sensors (labeled as S0...S7) were located in the plane of fMRI slice. The remaining sensors (S8...S14) were located in the slice half-way to the top, and the last channel (S15) on top of the head. EEG forward modeling was implemented using *OpenMEEG*[4] (see Figure S.9).

---
[3]TR of 5 sec was chosen as a reasonable tradeoff between conventional TR of 2-3 sec and long TR of 10.7316 sec in real data in Section 2.4.

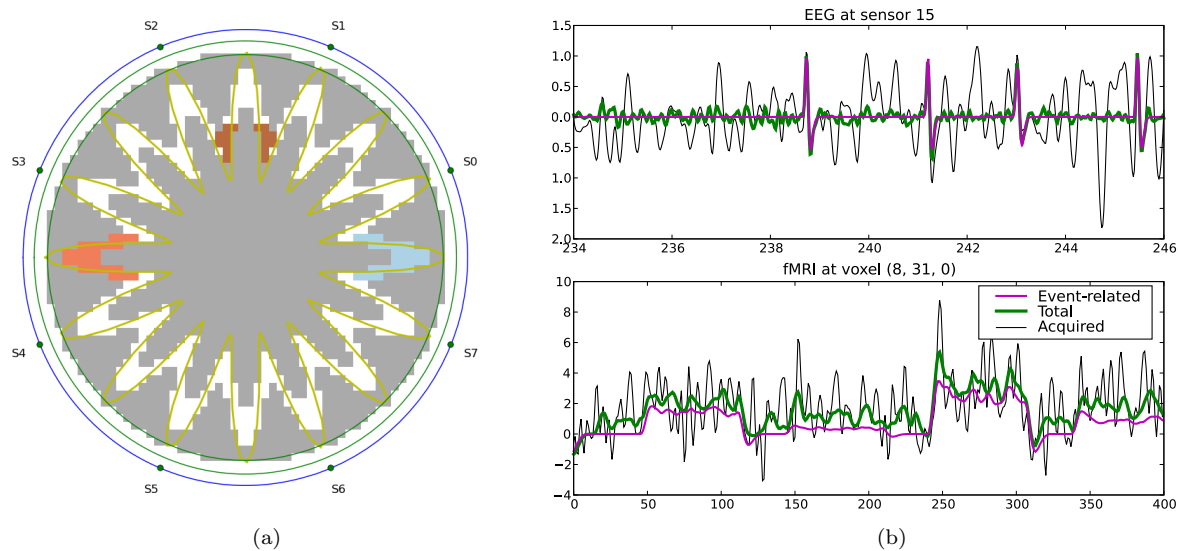

Figure 2: Simulation space and data. (a) The slice of "Spherical Brain": 2148 voxels with 3 ROIs to carry ER activity. (b) Samples of simulated EEG and fMRI signals. Event-related plots (in purple) depict simulated signal which is caused solely by ER activity. Total signal (in green) shows the total signal caused by all (ER and spontaneous) neural activity. The acquired data (in black) shows the total signal including additive noise.

Two types of neural activations were modeled – event-related (ER) and spurious activity. Event-related activity was added only within predefined ROIs. Onsets of ER activity were taken from the experimental design of the real dataset described in Section 2.4. Relative activation amplitudes in ROI1...ROI3 (see Figure 2(a)) were in the range from 0.5 to 1.3.

Since the brain is always active regardless of the actual stimulus, there is always neural activity unrelated to the extrinsic stimulus. This activity contributes variance to the acquired neural data that cannot be explained by the experimental design, and is different from the instrumental noise which should not have common cause in different data modalities. For the simulation, a range of different types of spurious activity were assigned randomly to different locations (see Figure S.7 for examples of the shapes of ERP responses). Amplitude and frequency of occurrences of spurious activity varied depending on the tissue-type. Within gray matter voxels spurious activity occurred with a frequency of 2 events per second and relative amplitude of 0.6. Within white matter voxels frequency and amplitude were half the size.

While modeling EEG responses, all simulated neural activity was jitterred uniformly within 50 msec of stimuli appearance to simulate the variability of the response per trial. Moreover, the response onset in each ROI was suject to an additional jitter (once again random within 50 msec) to model the response variability across brain regions. Both types of activations contributed to the total LFP signal assigned to each voxel. LFP signals of all brain voxels were transformed into EEG signals through the aforementioned forward EEG model. Jitterring and addition of spontaneous activity resulted in ERP responses, which looked very similar to the ones acquired empirically (see Figure 3(b)). Additional additive source of variance, instrumental noise, was modeled by random sampling of normal distribution with mean 0 and variance 1, with a consecutive low-pass Butterworth filtering[5] in temporal domain with stop-band frequency of 4 Hz (see Figure S.10).

---

[4] http://www-sop.inria.fr/odyssee/software/OpenMEEG

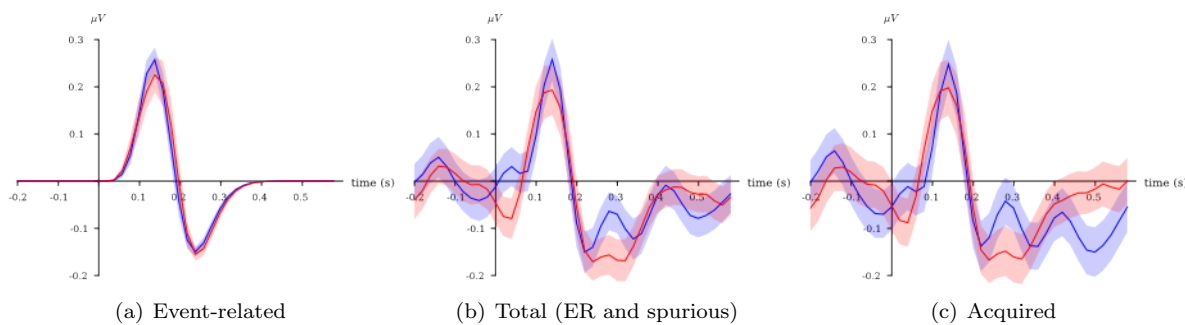

(a) Event-related  (b) Total (ER and spurious)  (c) Acquired

Figure 3: ERP of the simulated EEG data (red and blue colors distinguish two simulated levels of "activation"). (a) clean simulated even-related (ER) response with variance solely due to temporal jittering. (b) total activity – ER response with added spurious, not design-related activity. (c) Total activity with additive simulated instrumental noise.

Every fMRI voxel was modeled using a simple convolutional model of a linear time-invariant system. Squared values of the activation amplitudes were convolved with a conventional HRF (see Figure S.8). Additive instrumental noise was modeled similar to EEG noise, but the time with a 0.2 Hz stop-band frequency of the filter. Samples of the modeled signals are presented in Figure 2(b).

2.4. Real Data: Auditory Experiment

For this analysis data from a single participant in a previously published study (Thaerig et al., 2008) was taken. In that experiment subjects were exposed to monaural stimulation with linearly frequency modulated tones at two different sound intensities (60 dB and 80 dB). Experimental task was to detect the direction of frequency modulation of the tones. fMRI data was collected on a BRUKER 3T/60 head scanner using conventional FLASH sequence simultaneously with EEG recording using 29 sintered and nonmagnetic Ag/AgCl electrodes. fMRI data consisted of volumes with four slices (thickness 8 mm) acquired at TR=10.696 sec with an in-plane resolution of 3 mm. Further details about data acquisition setup, preprocessing stages, and the results of conventional analyses using GLM and ERP methods can be found in the original publication (Thaerig et al., 2008).

Due to a slight desynchronization of MRI and EEG acquisition equipments, the nominal fMRI TR of 10.696 sec was adjusted to 10.7316 sec to match the time clock of EEG acquisition hardware. This value was deduced by inspecting the EEG data carrying heavy artifacts that are caused by the MRI acquisition procedure (gradient sequences). The fMRI data were preprocessed using *FSL (Smith et al., 2004)*[6] tools: spatially smoothed at FWHM of 6 mm and temporally high-pass filtered at the cutoff corresponding to 140 sec. Only brain voxels were selected for the analysis.

The EEG data was initially preprocessed to remove MR-artifacts (see (Thaerig et al., 2008) for the details). Using *EEGLAB*[7] we have converted data from original proprietary format, temporally low-pass filtered at a cut-off frequency of 20 Hz, and downsampled to 50 Hz to match temporal resolution of the simulated data. All EEG data was re-referenced to the mean between TP9 and TP10 electrodes. 10 out of 29 EEG sensors (Cz, CP1, CP2, Pz, T7, T8, TP9, TP10, P7, P8) were selected to reduce the input dimensionality of the data. These sensors were selected as representative for providing a clean ERP response to the stimuli.

---

[5] Butterworth filter was chosen as the one providing frequency response as flat (minimal amount of "ripples") as possible.
[6] http://www.fmrib.ox.ac.uk/fsl
[7] http://sccn.ucsd.edu/eeglab

## 2.5. Data Preparation

EEG recordings of brain activity are dominated by the periods of rhythmic activity. It was suggested (Roopun et al., 2008) that characteristic bands of EEG signal, as generated by the neural matter, are centered at frequencies arranged according to the powers of "golden ratio" $\phi = 1.61803399$. Thus, prior to the analysis, the EEG signal of each channel was transformed into its power spectra, and was summarized by the total power in each temporal/spectral bin, taking $\tau = 10$ sec (see (3)) prior to the onset of each fMRI volume. The temporal width of each bin was taken to be 500 ms and frequencies followed the aforementioned $\phi$ distance in a log space to cover the range up to 25 Hz available within the simulated signal sampled at 50 Hz. That resulted in 20(time-bins) × 9(spectral-bins) × 16(electrodes) = 2880 features in the simulated EEG. The number of samples matched the number of simulated fMRI volumes. Each EEG feature, and each fMRI voxel were independently standardized (Z-scored) prior to the analysis.

## 2.6. Data Analysis

The TRANSfusion analysis was implemented using *PyMVPA (Hanke et al., 2009)*[8], which provides a convenient interface to *LIBSVM's*[9] SVR and facilitated the construction of pipelines for the analysis of neuroimaging data using statistical learning methods. Complete simulation and analysis code, and EEG/fMRI data used in the paper available from http://www.pymvpa.org/files/papers/fusion2010-1. The provided analysis script implements data import, spectral decomposition and binning, basic feature preparation (detrending and zscoring), standard 4-fold cross-validation per each fMRI voxel using linear SVR on noisy data (EEG and fMRI), visualization and storage of results.

## 3. Results

### 3.1. Simulated Data

The reconstruction performance was characterized by a correlation coefficient between the target (not observed during training) and predicted timeseries. The reconstruction of noisy unseen data across all voxels of the brain is presented in Figure 4(a). The figure only shows voxels which passed thresholding at z=3.0 relative to chance performance. This should result in only 0.1% false positives in the results (*i.e.*, on average 2 voxels from the respective population). To assess the by-chance performance distribution, thresholding relied on the assumption that the performance histogram is a mixture of two distributions: a normal distribution[10] centered at 0 which corresponds to by-chance reconstruction performance, and the other, arbitrarily shaped, distribution of valid reconstruction performances, which does not have any significant number of negative values. This assumption readily allows to assess the standard deviation of the distribution for by-chance performance simply by mirroring negative performance values around zero and computing the standard deviation of the obtained distribution. Such strategy was also used for the analysis of real data in Section 3.2.2. Alternatively, non-parametric testing could be done in a similar fashion if the distribution is assumed to be symmetric but not necessarily normally distributed.

When an analysis is done on simulated data, it is possible to measure the performance by inspecting the reconstruction relative to the noiseless data. Figure 4(b) shows that the reconstruction performance is significantly higher when compared to the noiseless data, and that the distribution of by-chance performances is visibly segregated from the distribution of the meaningful reconstructions.

Results on the simulated data allowed to conclude that TRANSfusion primarily allows to detect areas with relatively strong (not spurious) neural activity. We argue that spurious activity, which is asynchronous among voxels, is not well represented in the captured EEG signal, and hence does not help in reconstructing corresponding voxel time-courses. Another interesting observation is that although EEG signal was simulated using templates of ERPs and not oscillatory EEG sources, and TRANSfusion relied on actual power-spectrum representation of the signal, it was, nevertheless, able to reconstruct significant variance of simulated BOLD fMRI responses. Therefore, TRANSfusion not only seems to be effective for the analysis of EEG signals with good localization in spectral domain, but also for signals with typical shapes of ERP responses.

---

[8] http://www.pymvpa.org
[9] http://www.csie.ntu.edu.tw/~cjlin/libsvm
[10] This was actually an *rdist*[11] distribution, which is well approximated by a normal distribution for the relatively large sample sizes.

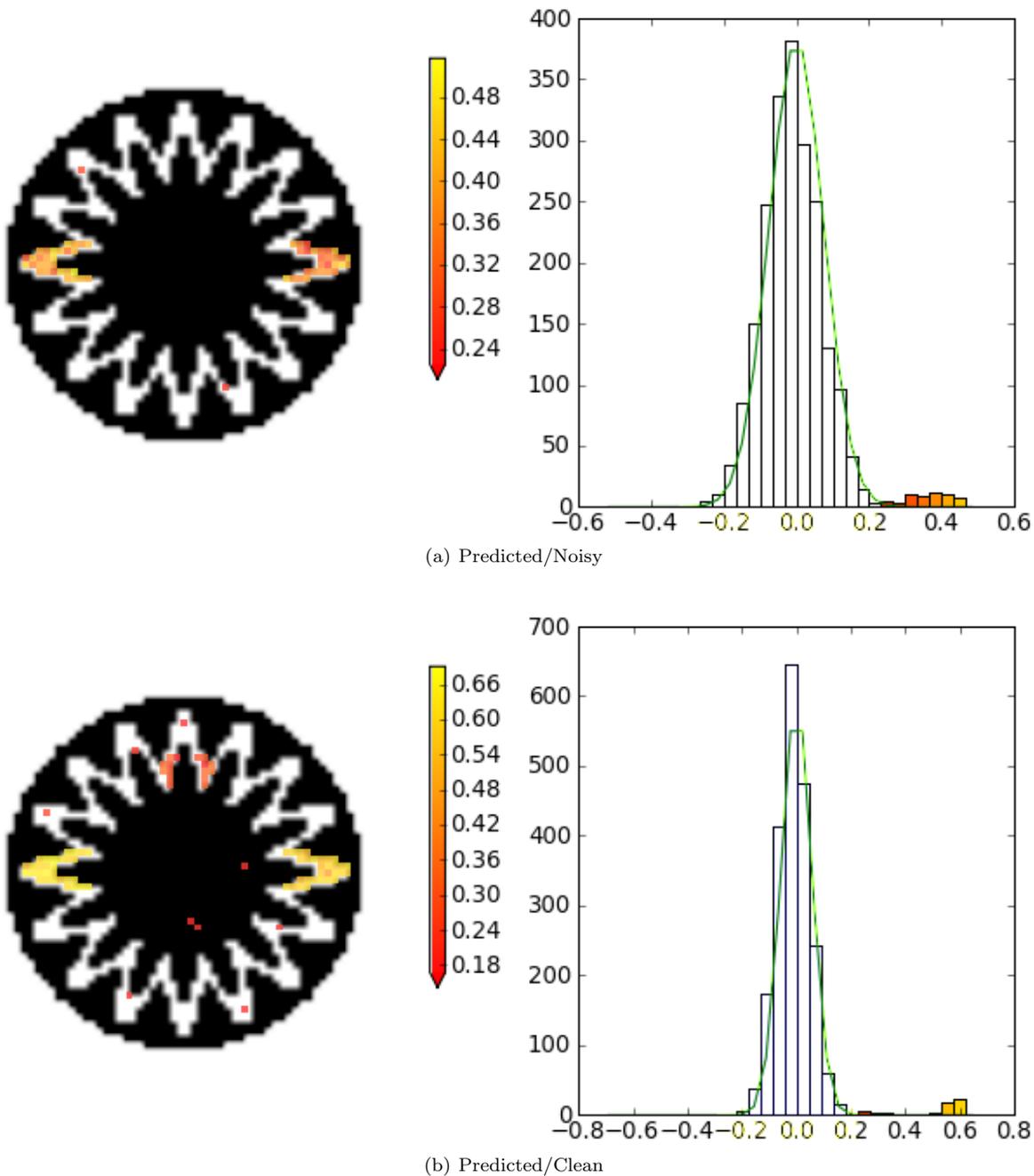

Figure 4: Thresholded correlation coefficients of predictions with noisy and clean original fMRI data. Plots on the right: histograms of correlation coefficients for predictions, with blank bars corresponding to values which have not passed the threshold of z=3.0 (corresponding to $p \approx 0.001$), and suprathresholded values colored according to the color bar in the middle. Chance performance distribution (in green) was estimated by mirroring negative performance values around the zero point and fitting normal distribution to obtained histogram for the purpose of statistical testing. Plots on the left: spatial maps of locations with supraphresholded estimates.

### 3.2. Real Data

#### 3.2.1. Signal Reconstruction

The analysis workflow was the same as for the simulated data in Section 3.1 with a single difference. In this dataset the TR was relatively long (10 sec), which means that the actual acquisition of the slices was spread through that duration. To not rely on the sequence of slice acquisition, $\tau = 20$ sec was taken to integrate a sufficient amount of information to account for the fMRI data in all four slices.

Reconstruction performance of the data was statistically significant across a variety of brain regions (Figure 5(a)). These results are surprising in the sense that the reconstruction achieved relatively high performance not only in areas located in the vicinity of EEG electrodes, but also in medial areas which, presumably, should be weakly represented in EEG signal registered on the scalp. This effect might be attributed to strong activations in those areas while subjects were performing the task. Indeed, all areas detected by the GLM analysis (Figure 5(b)) are also present in the thresholded reconstruction map (Figure 5(a)). Additionally, left auditory cortex which was not significantly active according to the GLM results, was reconstructed at a comparable level as the right auditory cortex. Possible explanation could be due to laterality in auditory processing. It is known that right auditory cortex is dominantly activated in the tasks concerned with discriminating tonal (frequency) differences of sound patterns, whereas left auditory cortex was found to be engaged primarily in processing of temporal structure of the auditory stimuli (*e.g.*, Zatorre and Belin, 2001), thus activity specific for language processing is often attributed to the left auditory cortex (Näätänen et al., 1997). In the given experiment stimuli were linearly frequency modulated, but GLM model was not accounting for the effects of upward or downward modulation happening through the temporal evolutions of the trials. Therefore, GLM modeling could not account for such source of variance which lead only to detection of consistent through the trials activations in the right auditory cortex.

#### 3.2.2. Sensitivity Analysis

Many multimodal studies (*e.g.*, Goldman et al., 2002; Laufs et al., 2003; Martinez-Montes et al., 2004; de Munck et al., 2007) aim at localizing brain regions which are generators of a specific variation (*e.g.*, frequency band) of cortical activity. The analysis of SVR sensitivities to different spectral bands allows to detect areas where a particular frequency band contributes most of the variance toward the reconstruction of fMRI BOLD signal. Figure 6(a) shows thresholded ($z > 2.0$ under assumption of normally distributed values) aggregate sensitivities to the EEG signal spectral components in the $\alpha$-band. Highest sensitivities are found in the occipital areas. This finding is consitent with previous studies (Goldman et al., 2002; Goncalves et al., 2005). Unfortunately, the available fMRI data did not cover the full brain, and therefore made it impossible to explore all brain regions in regards to the localization of generators of different frequency bands (Figures S.18, S.19, and S.20).

Another possible dimension for the sensitivities aggregation are sensitivity plots of voxels per each EEG sensor. Figure 6(b) shows thresholded aggregated sensitivities of different voxels to the signal of the T8 EEG sensor. Results indicate that information from T8 sensor describes primarily activity from right posterior regions of the brain. Nevertheless, some more distant areas are also sensitive to the data in T8. For instance, parietal regions, whose activation pattern is, according to GLM results (see Figure 5(b)), not experiment-related, nevertheless, carry BOLD signal which is well described by EEG (see Figure 5(a)). Whether those regions are just weakly reflected in the T8 channel or actually activate coherently with auditory cortex remains an open question. Since it is unlikely for a T8 electrode to pick up significant signal from distant areas in the contra-lateral hemisphere, one could also speculate that contra-lateral supra-thresholded voxels are the entry points to the cortical structure from the corpus collosum which facilitates the communication between two hemispheres. Hence such simple analysis of sensitivities might allow to deduce functional connectivity between distant areas. Aggregate sensitivities of other electrodes also seems to be in accordance with their spacial locations. For example, significant sensitivities of Pz, CP1, and CP2 electrodes cover parietal areas (Figures S.21, S.22, S.23).

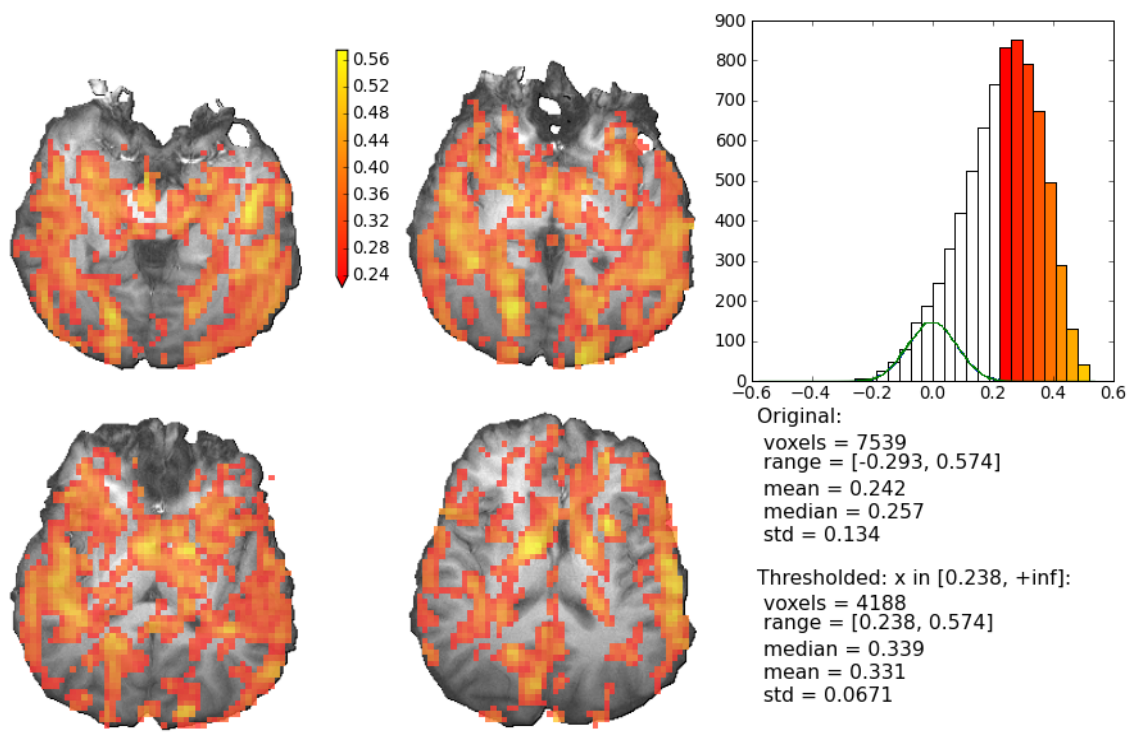

(a) TRANSfusion

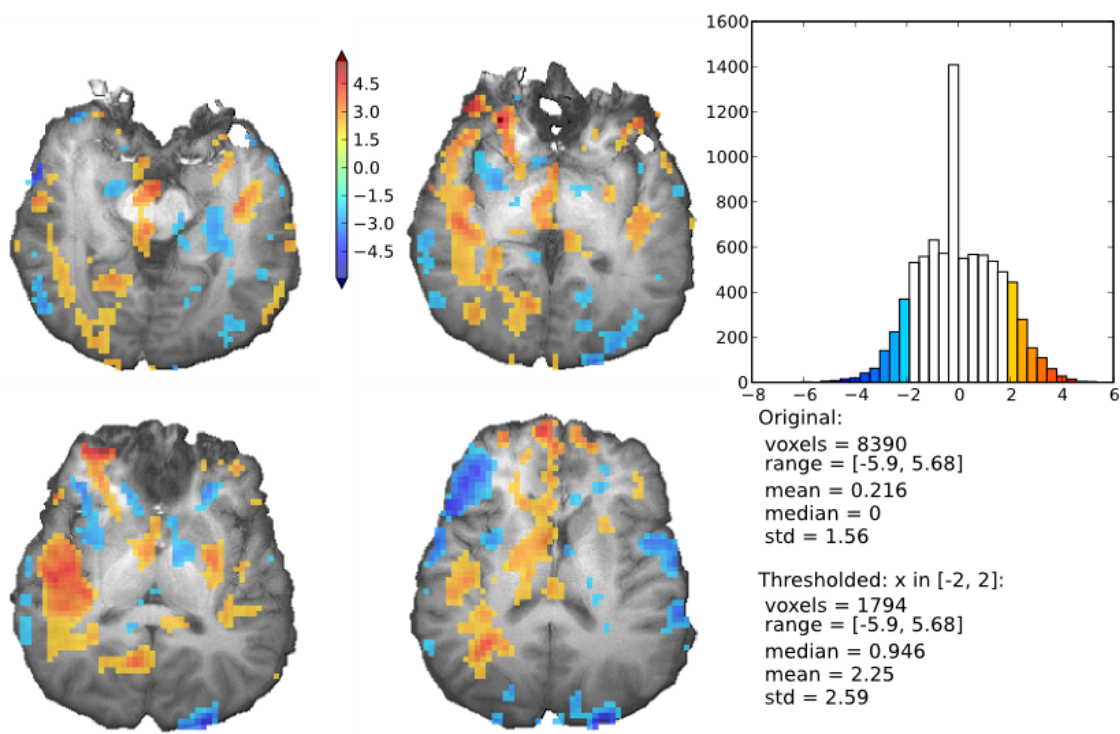

(b) GLM

Figure 5: Activated areas as detected by TRANSfusion vs. fMRI GLM results. Slices are plotted in radiological convention, hence left side of the brain is presented to the right from inter-hemispheric point. (a) Thresholded correlation coefficients between real and predicted fMRI data. (b) Significant (thresholded at $|z| = 2.0$) activation (in red) and deactivation (in blue) in response to auditory stimulation.

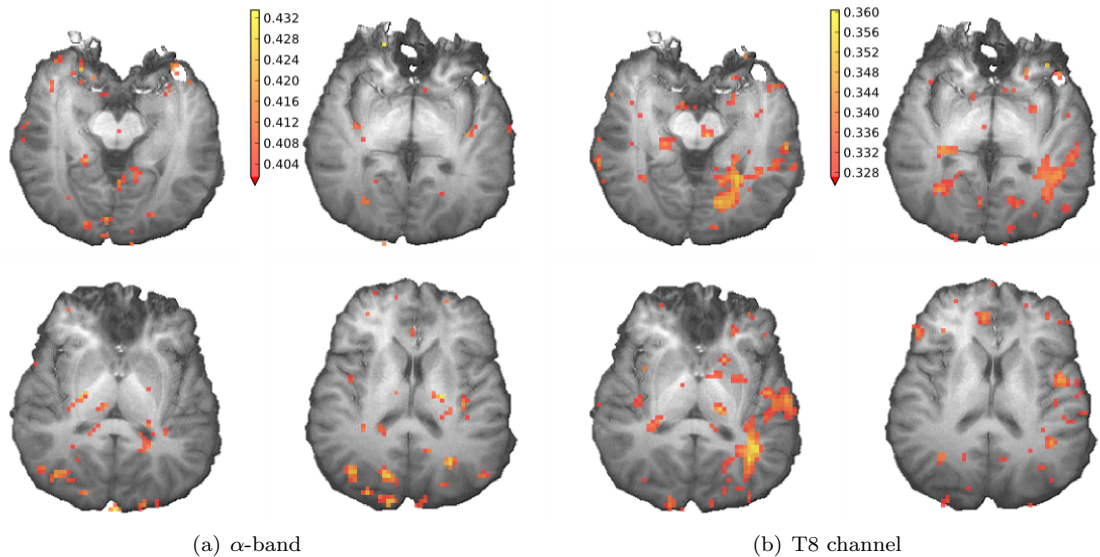

(a) $\alpha$-band        (b) T8 channel

Figure 6: Aggregate sensitivities to (a) frequency band and (b) an EEG sensor.

## 4. Discussion

For every acquisition method capturing reflections of neural activity specialized techniques have been developed to segregate actual neural signal from various types of additional variance (*e.g.*, instrumental and physiological noise). At a gross-level, all of the existing instrumental methods capture the same activity spatially and/or temporally filtered, quantized at different resolutions, and possibly subject to some non-linear transformation. Despite the loss of some information due to imaging modality specifics (*e.g.*, spatial specificity in EEG), a considerable amount of the same neural processes should be reflected in any modality capable of full-brain capture.

Therefore, in addition to the *fusion* approach, where the ultimate goal is to combine multiple modalities into a single meta-modality carrying the best aspects of all, the less ambitious TRANSfusion approach might be worth considering. TRANSfusion aims at modeling the signal of one modality in terms of another. By doing so it cannot really reach the ultimate goal of providing a meta-modality with superior spatial and temporal specificity. Nevertheless it can be useful for performing activity localization, exploring data for coherence/connectivity structure, and even boosting temporal resolution (not necessarily specificity) of fMRI.

In this paper we have demonstrated the application of TRANSfusion to simulated and real data by creating a reliable mapping from EEG onto fMRI data. Analysis of this mapping allowed to identify brain regions, where a proportion of the fMRI signal can be reliably explained by the information from EEG. These results not only provide localization of the ongoing neural activity, but also reveal interconnectivity among the areas. Since deep structures are presumably weakly represented in EEG, the goodness of their description in terms of EEG signal relies on their connections with more exterior parts of the brain. The analysis of coherence with external structures was further elaborated by the analysis of sensitivity maps across brain regions per each EEG channel. Further analysis of sensitivity to different components of EEG signal allowed to localize possible generators of corresponding components, and/or regions where neural activity was modulated by them.

Besides conventionally explored frequency bands of EEG, very low-frequencies ($f < 0.1\,\text{Hz}$, also known as DC-EEG) signal component has not yet been a subject of attention for multimodal integration despite recent experiments showing the strong correlation between the changes of the observed DC-EEG signal and the hemodynamic changes in the human brain (Vanhatalo et al., 2003). In fact, such DC-EEG/BOLD coupling suggests that the integration of fMRI and DC-EEG might be a particularly useful way to study the nature of the time variations in hemodynamic signal. These variations are usually observed during fMRI experiments but are not explicitly explained by the experimental design or by the physics of the MR acquisition process. Preliminary results in the sensitivity profiles across different frequency bands (see Figure S.13) suggest that those low frequencies play significant role in the description of fMRI BOLD signal.

The TRANSfusion approach does not impose any strict forward model for the BOLD response, it is capable of accounting for the sources of variance present in multiple (or all) EEG channels, and can be easily extended for non-linear mappings between EEG and fMRI. These aspects make this methodology versatile and actually applicable to real data.

In addition to applications of TRANSfusion presented in this paper, this technique could be natively extended to cover a range of additional applications. For instance, learnt model provides basis for EEG and fMRI filtering based on the results of multimodal mapping; assessment of descriptive power for any given electrode (or ICA component) provides grounds for electrodes (component) selection; learnt model could be readily applied to provide interpolation (or even extrapolation) and thus improved temporal sampling of fMRI signal based on EEG signal taken at denser temporal resolution which is available in EEG but is not provided by MRI equipment. This reliable interpolation of fMRI might be beneficial for slice-timing correction and might also be beneficial for causality graph discovery where timing might be of critical importance. Furthermore, contrasting (as simple as paired t-test) reconstruction maps obtained using EEG/fMRI data from different conditions would allow to localize differentially engaged areas (once again without explicit reliance on any HRF model). We argue that since it does not require the specification of an experimental design, TRANSfusion approach might be beneficial for the analysis of data with spontaneous (*e.g.*, epileptic) activity, or ongoing activity (*e.g.*, resting state), where it is impossible to (pre-)define variable or locations of interest.

*Acknowledgments*

Dr. Yaroslav O. Halchenko were supported by the National Science Foundation (grant: SBE 0751008), James McDonnell Foundation (grant: 220020127), and XXX.

# Supplemental Materials

## Appendix A. Simulated Data

### Appendix A.1. Simulated Signals

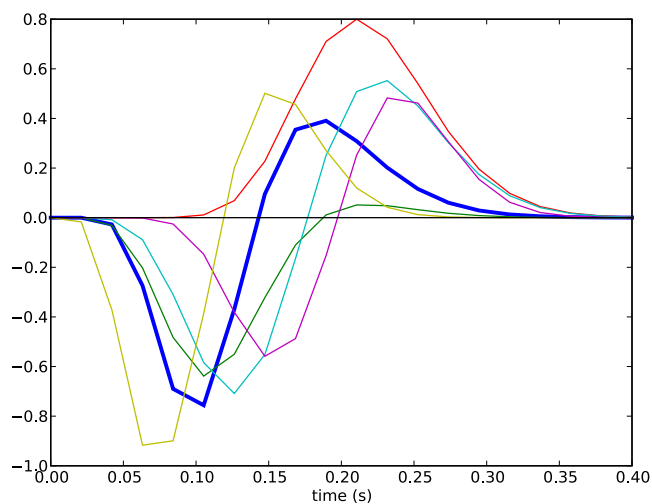

Figure S.7: LFP responses used for simulation. Blue LFP is the one used for modeling event-related activity, while all the others (including the blue one as well) were arbitrarily assigned to voxels to generate simulated LFP response.

### Appendix A.2. Analysis Results

To visually inspect the goodness of reconstruction, time series for voxels from four location are presented in Figure S.11, which shows that time series within ROI1 were relatively well reconstructed, especially in respect to the clean original signal, although the learner was only exposed to the noisy EEG and fMRI data during training.

### Appendix A.3. Sensitivity Analysis

Demonstrated reconstruction of the original time-series within the ROI voxels allows to state, that the regression has learned information about neural activation which was embedded in both EEG and fMRI signals. It allows further to inspect the sensitivities of the trained regression. Comparison of sensitivities across all sensors (see Figure S.13) shows that results in the sample voxels within ROI1 are highly sensitive to the high frequencies components of the S0 (Figure S.12).

Aggregate sensitivity (sum of powers of the sensitivities) in the range of high-frequencies for that sensor is presented in Figure S.12. Maximum sensitivity nicely matches the time-lag of 5 sec where HRF used for fMRI simulation reaches its maximum value (see Figure S.8). Absence of the sensitivity to the lower frequencies of EEG signal could be attributed to the fact that simulated instrumental noise added to EEG signal was low-pass filtered, hence contributing mostly to the lower frequencies, thus making them less reliable for the estimation of the BOLD response. Obtained results allow to conclude that the analysis of the regression sensitivities provides means to assess characteristics of the underlying HRF of the BOLD response.

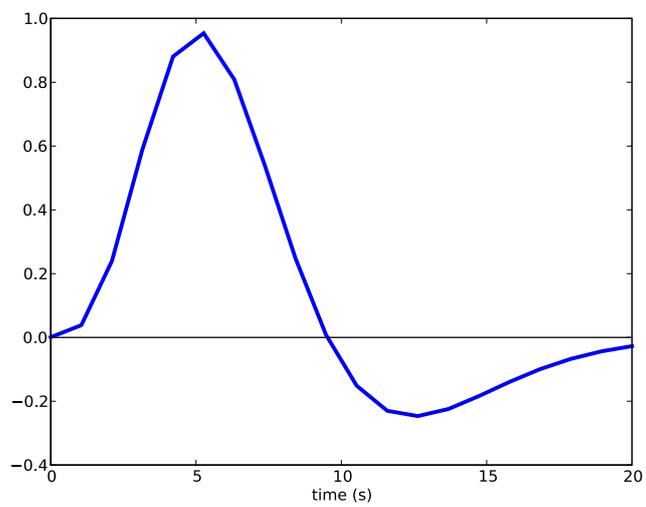

Figure S.8: HRF response used for fMRI data simulation.

## Appendix B. Real EEG/fMRI Data

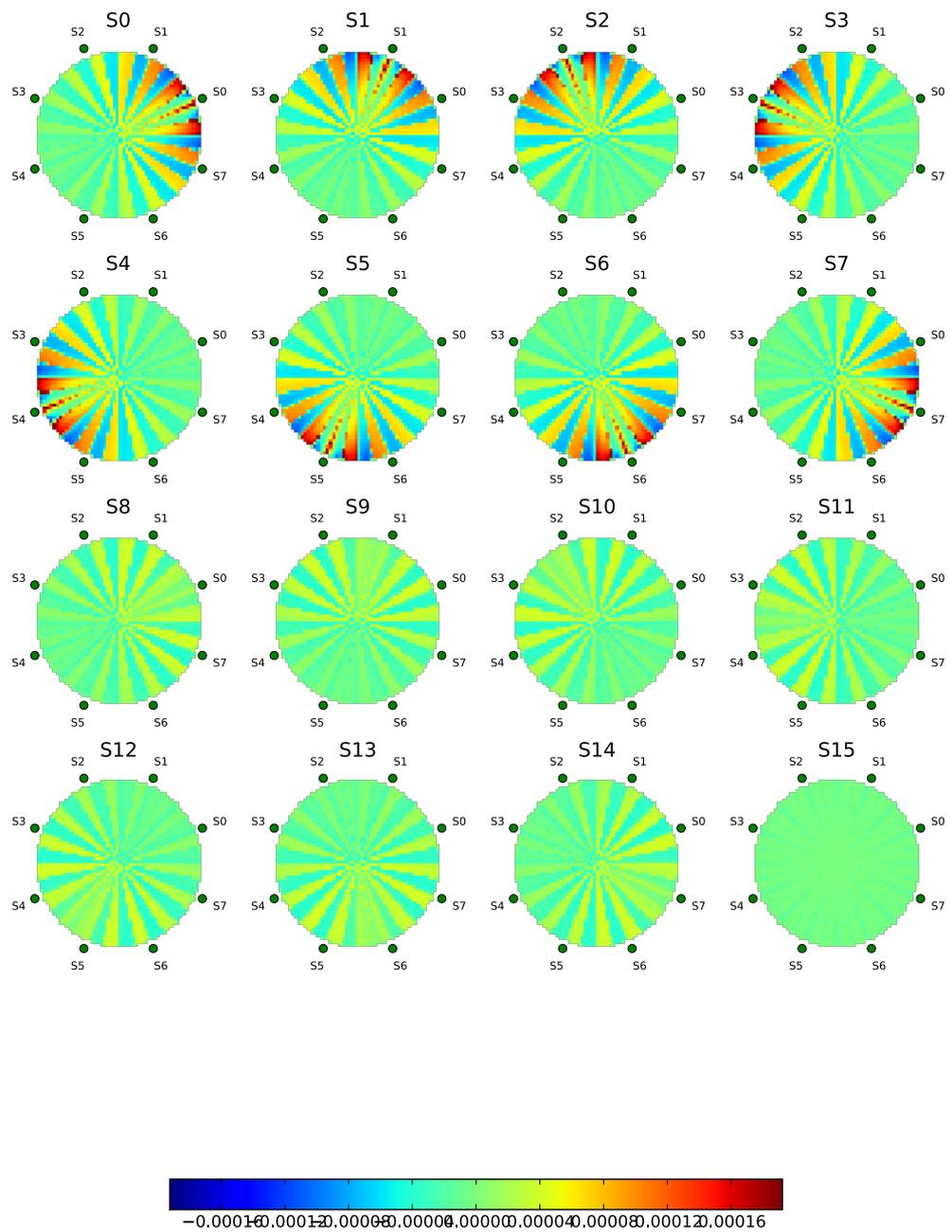

Figure S.9: Gain matrix for EEG signal simulation.

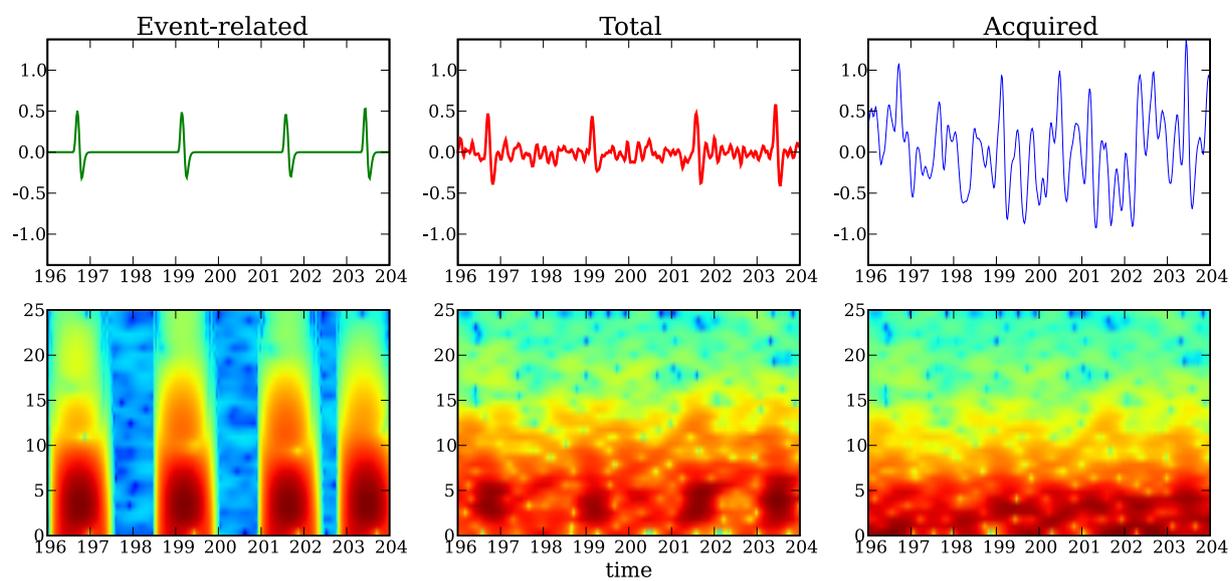

Figure S.10: Spectral characteristics of simulated EEG signals. The upper part presents sample durations of EEG. The lower part shows corresponding spectrograms.

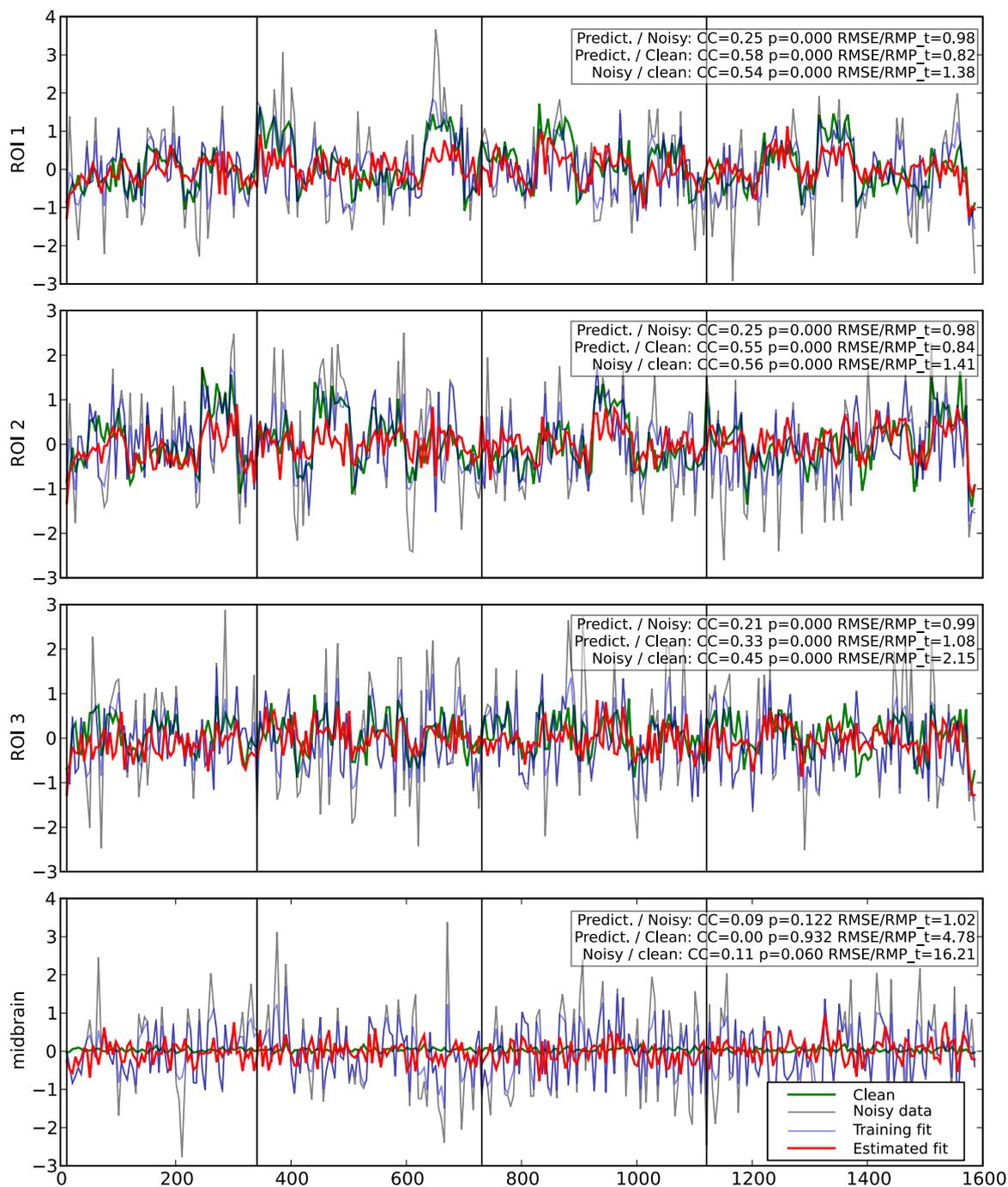

Figure S.11: Original and predicted fMRI time series for representative voxels. CC stands for correlation coefficient (or Pearson's correlation) with a corresponding significance level $p$. RMSE/RMS_t is the ratio between root-mean squared error (RMSE) and root-mean square of the target signal (RMS_t). Perfect reconstruction would correspond to CC=1.0 and RMSE/RMS_t=0.0. While voxels in the active areas are relatively well reconstructed, midbrain inactive voxel is not reconstructed at all.

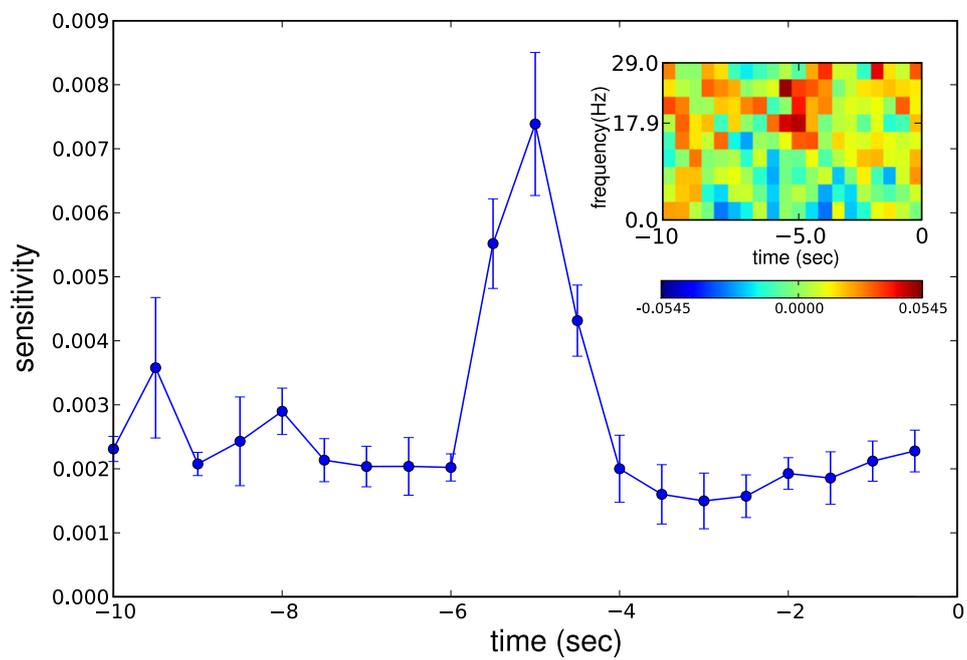

Figure S.12: Aggregate sensitivity of SVR to the high-frequencies of S0 EEG electrode while predicting a neighboring voxel in ROI1. Error-bars correspond to the standard error estimate across 4 splits of data. Embedded window shows sensitivities of S0 electrode to all time/frequency bins.

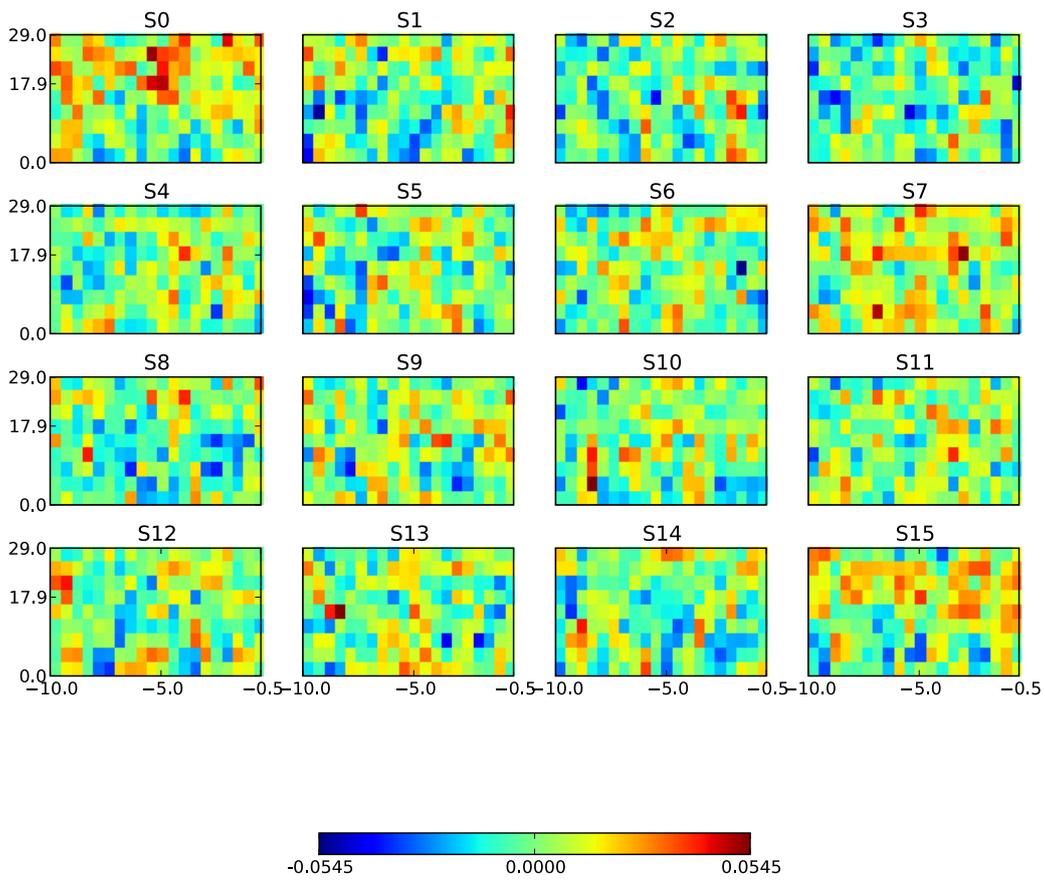

Figure S.13: Sensitivities of SVR trained to predict fMRI signal of a voxel in ROI 1. Each plot corresponds to a single channel of EEG and represents sensitivities in time (horizontal) and frequency (vertical) bins.

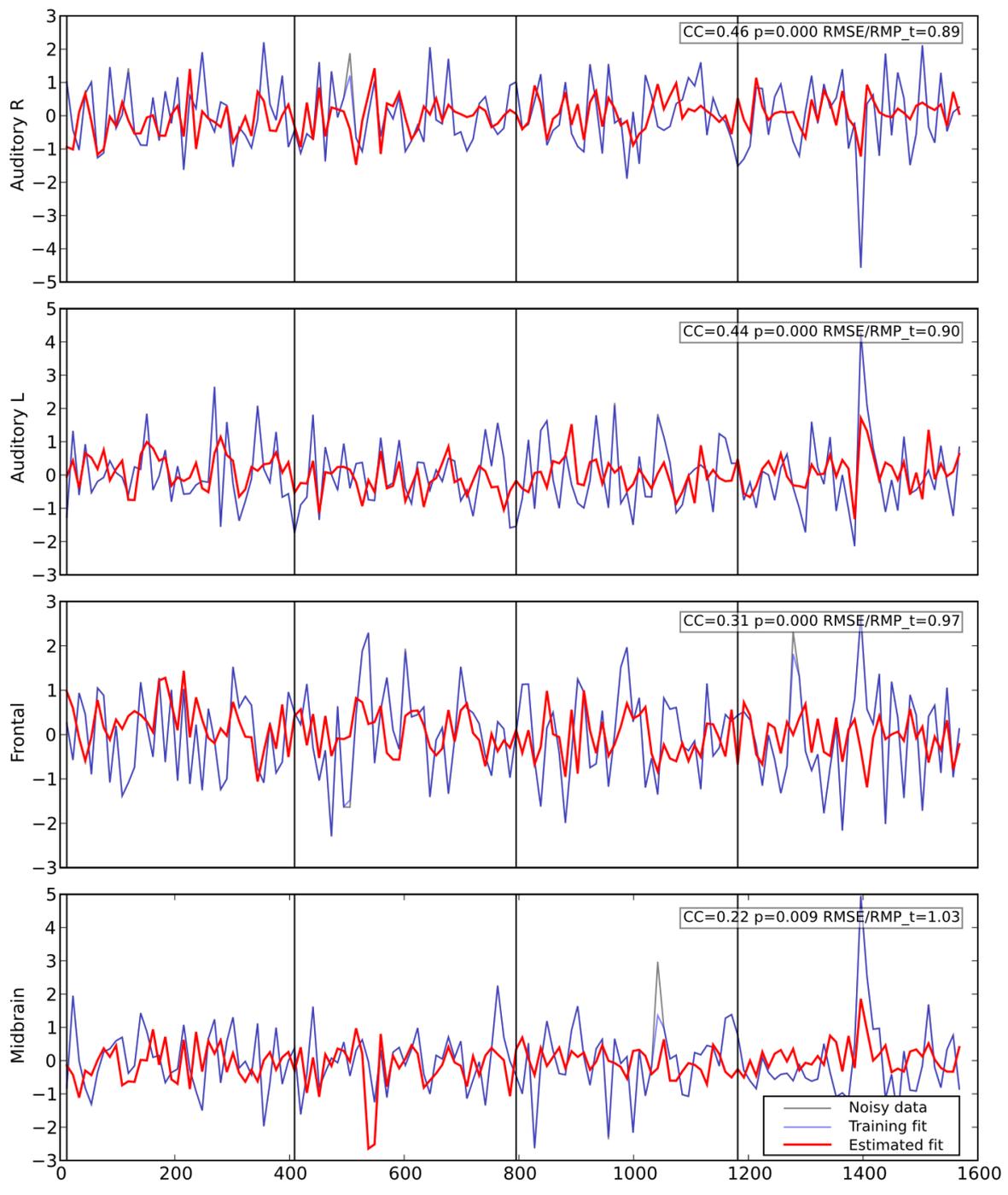

Figure S.14: Original and predicted fMRI time series for representative voxels.

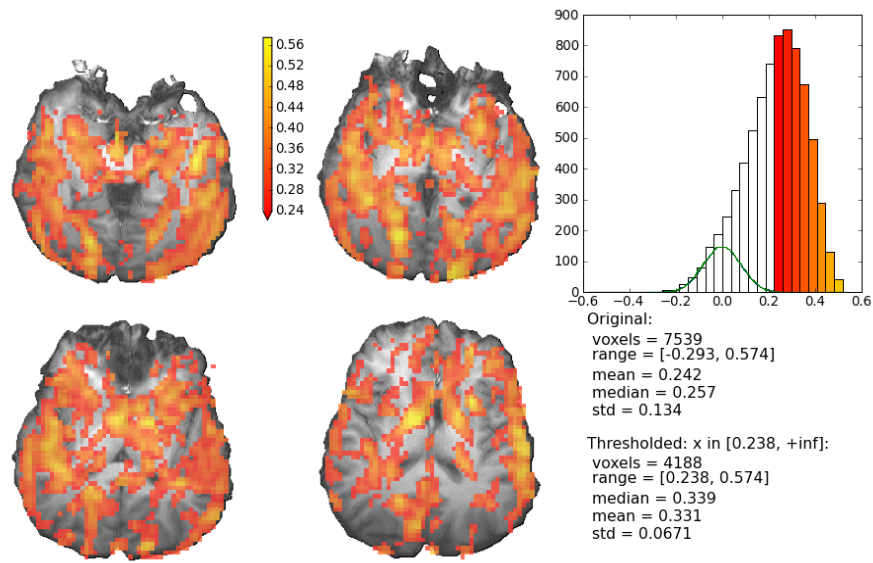

Figure S.15: Thresholded correlation coefficients between real and predicted fMRI data in a single subject.

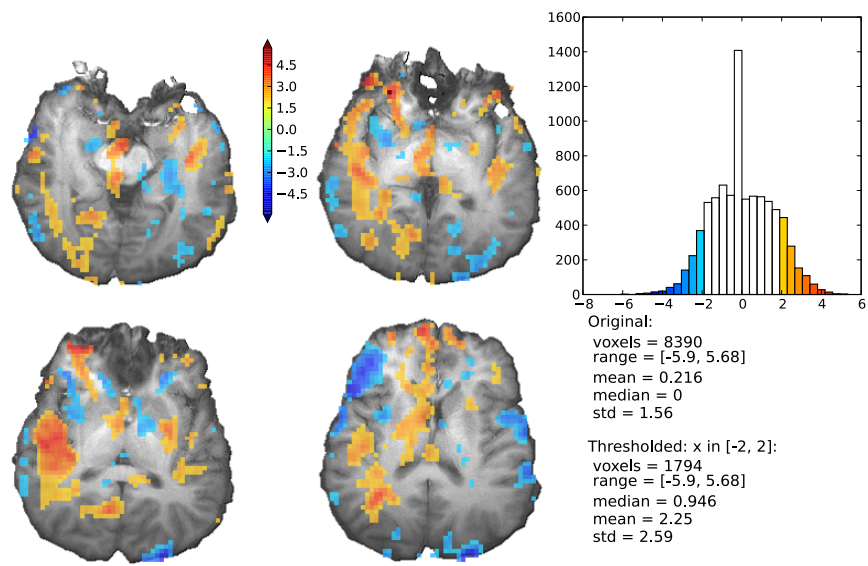

Figure S.16: GLM analysis: significant (thresholded at $|z| = 2.0$) activation (in red) and deactivation (in blue) in response to auditory stimulation.

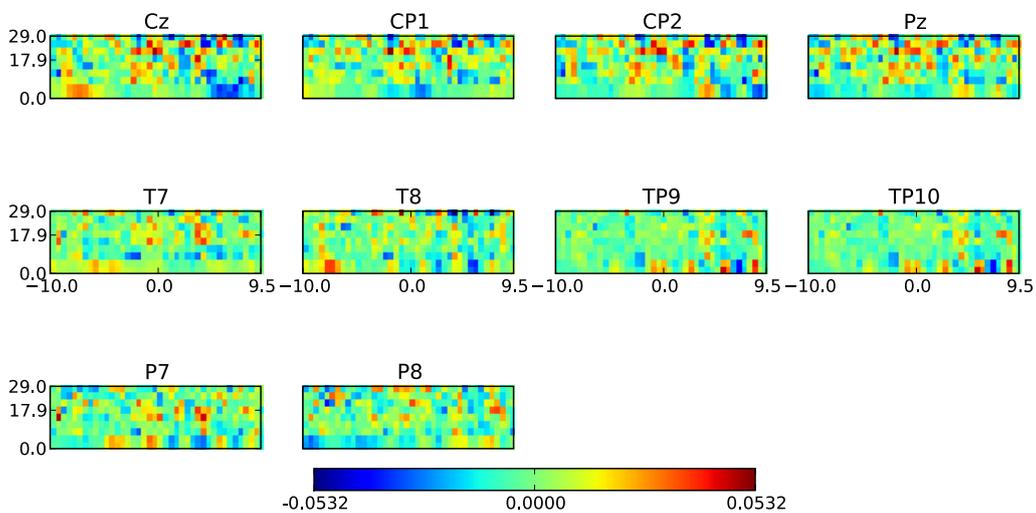

Figure S.17: Sensitivities of SVR trained to predict fMRI signal of a voxel with right auditory cortex. Each plot corresponds to a single channel of EEG and represents sensitivities in time (horizontal) and frequency (vertical) bins.

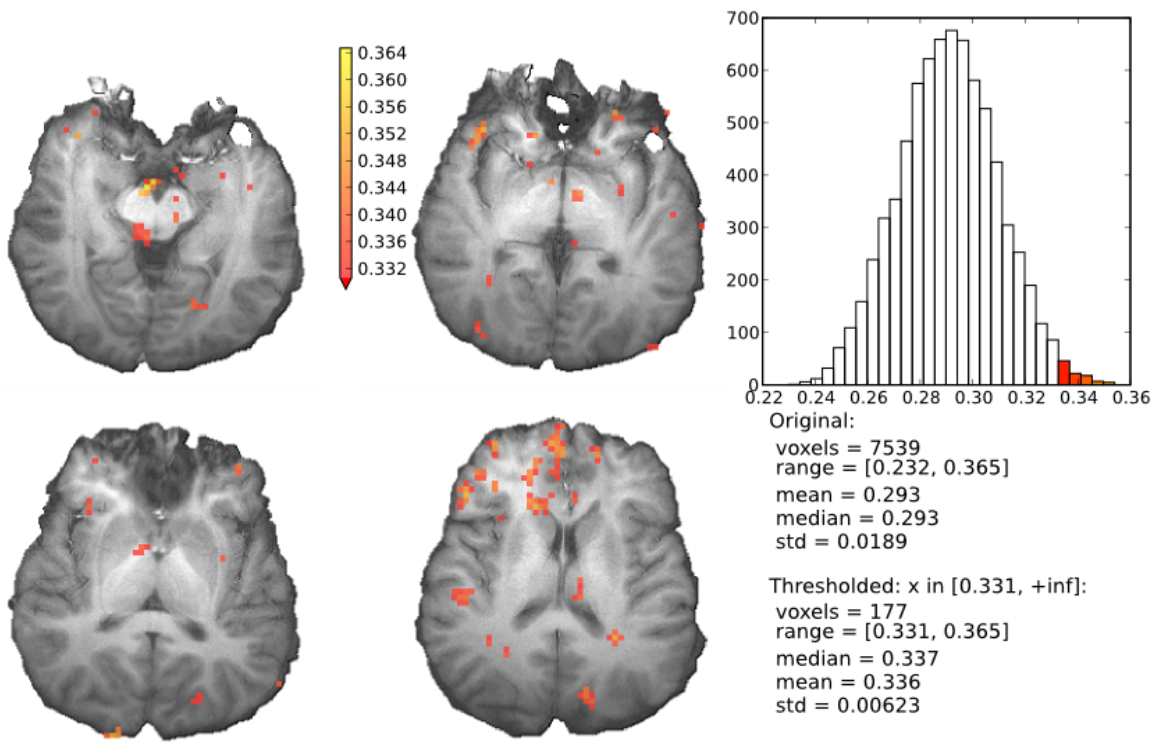

Figure S.18: Aggregate sensitivity to the frequencies in the $\delta_1$-band range of frequencies.

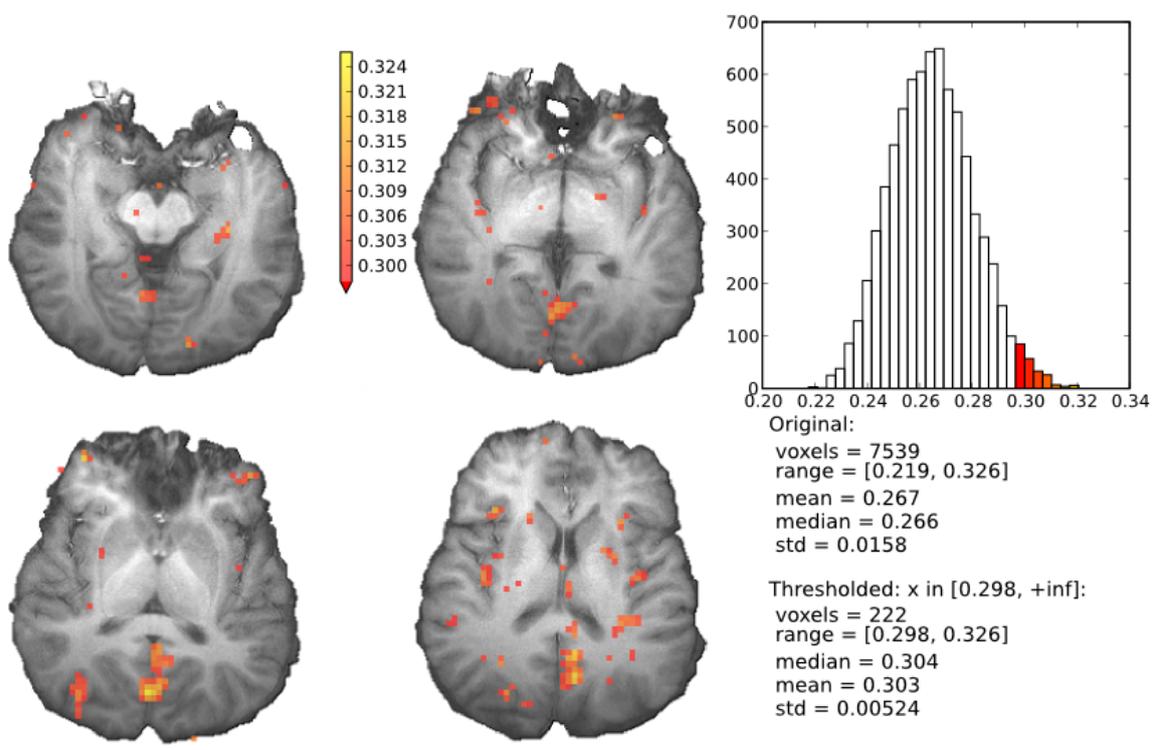

Figure S.19: Aggregate sensitivity to the frequencies in the $\delta_2$-band range of frequencies.

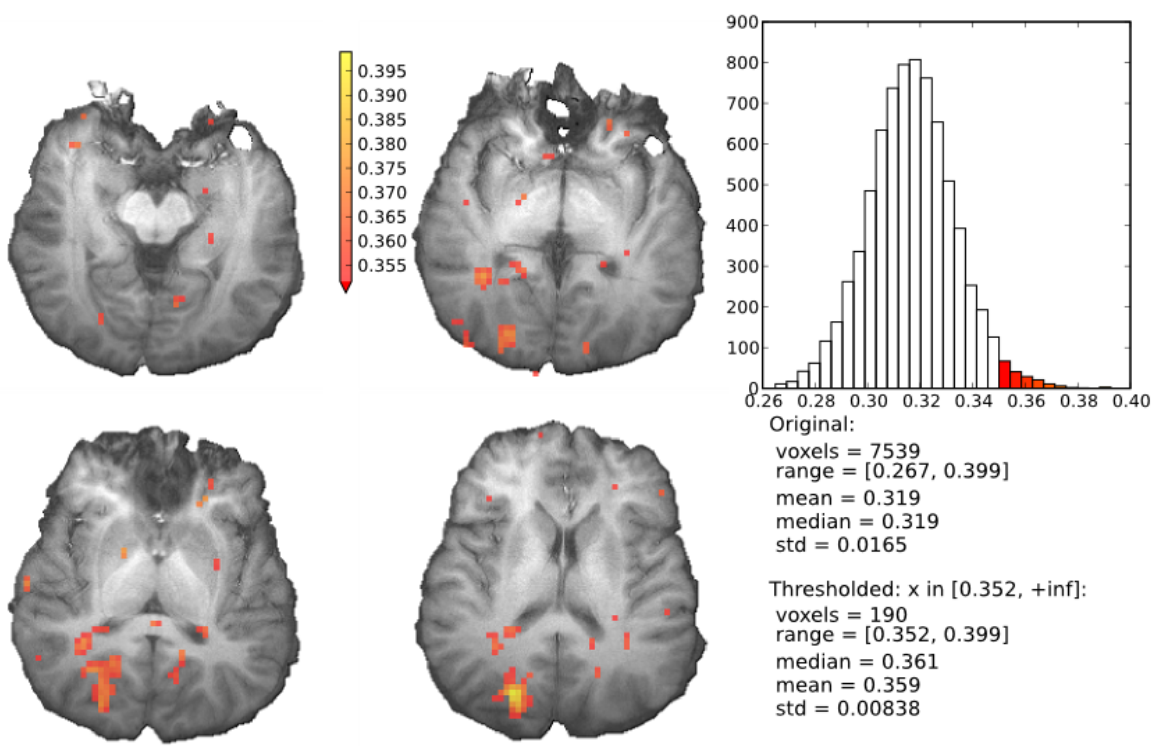

Figure S.20: Aggregate sensitivity to the frequencies in the $\theta$-band range of frequencies.

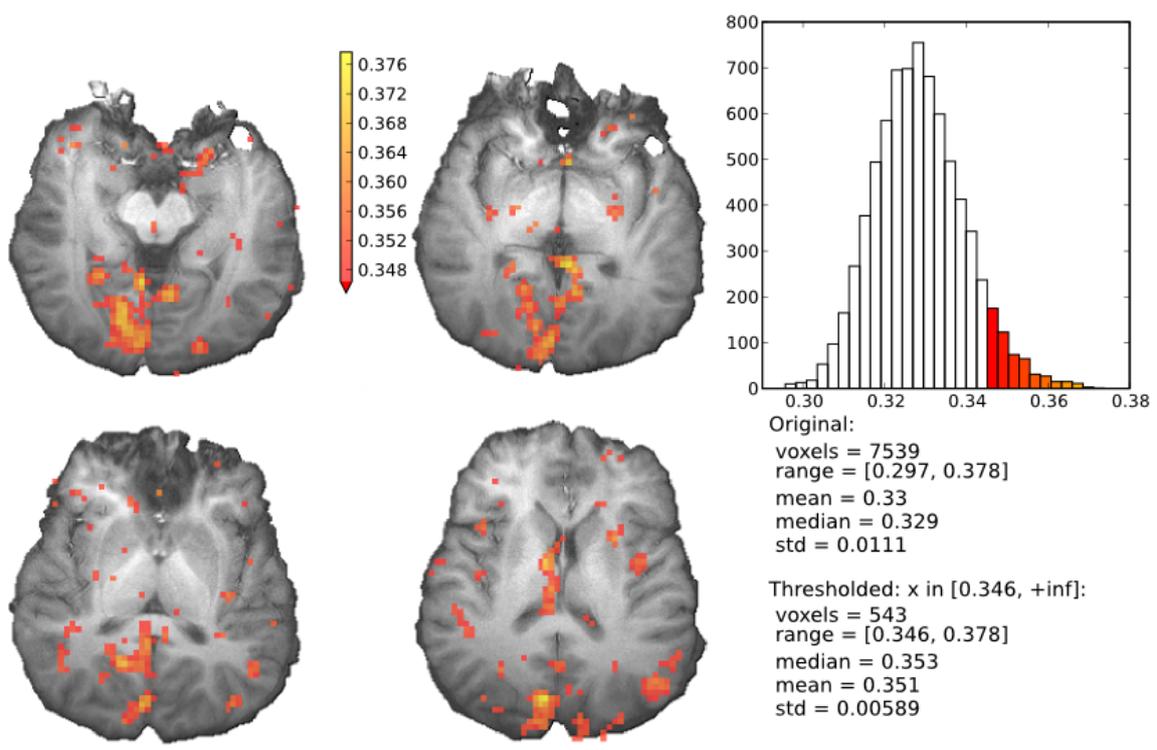

Figure S.21: Aggregate sensitivity to the signal of CP1 EEG channel.

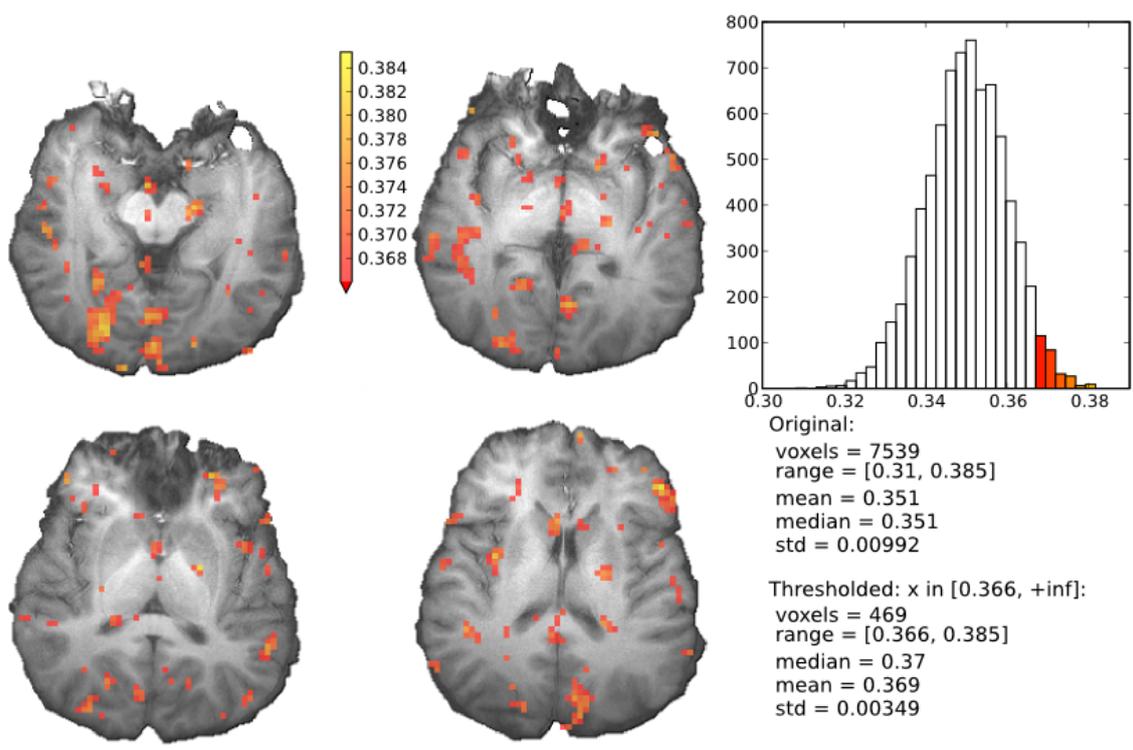

Figure S.22: Aggregate sensitivity to the signal of CP2 EEG channel.

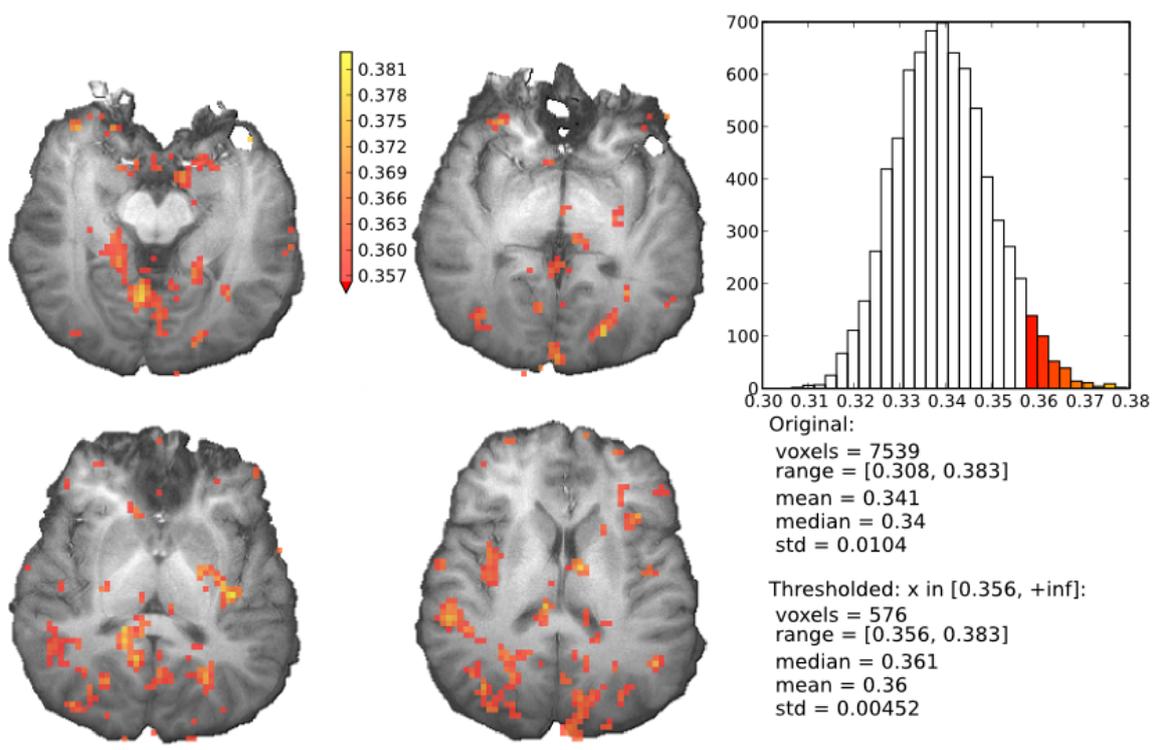

Figure S.23: Aggregate sensitivity to the signal of Pz EEG channel.